\documentclass[11pt, a4paper, logo, onecolumn, external, copyright]{pacificfusion}

\usepackage{amssymb}
\usepackage{color}
\usepackage[sort&compress,numbers]{natbib}
\usepackage{xcolor}
\usepackage{appendix}
\usepackage{subcaption}
\usepackage{mwe}
\usepackage{graphicx}
\usepackage{xr}
\usepackage{marvosym}

\graphicspath{{assets/images/}}

\begin{document}

\title{Opportunities in Pulsed Magnetic Fusion Energy}

\author[1]{C. Leland Ellison}
\author[2]{Vincent Garcia}
\author[3]{Matthew Gomez}
\author[4]{Gary P. Grim}
\author[4]{Jim H. Hammer}
\author[3]{Christopher A. Jennings}
\author[1]{Patrick Knapp}
\author[1]{Keith R. LeChien}
\author[1]{Nathan Meezan}
\author[2]{Robert Peterson}
\author[5]{Adam Reyes}
\author[3]{Adam Steiner}
\author[4]{William A. Stygar}
\author[5]{Petros Tzeferacos}
\author[6]{Dale Welch}
\author[1]{Alex Zylstra}

\affil[1]{Pacific Fusion, Fremont CA, USA}
\affil[2]{Los Alamos National Laboratory, Los Alamos NM, USA}%
\affil[3]{Sandia National Laboratories, Albuquerque NM, USA}
\affil[4]{Lawrence Livermore National Laboratory, Livermore CA, USA}
\affil[5]{University of Rochester, Rochester NY, USA}
\affil[6]{Voss Scientific, Albuquerque NM, USA}

\renewcommand{\today}{18 July 2024}
\date{\today}

\begin{abstract}    
    Fusion is a potentially transformational energy technology, which promises limitless clean energy. Yet, it requires continued scientific and technological development to realize its potential. The conditions necessary for fusion energy gain in terms of the product of plasma pressure $P$ and confinement time $\tau$ have been known for many decades. An underappreciated fact is that pulsed magnetic fusion has demonstrated $P \tau$ performance on par with laser-driven ICF and tokamaks despite receiving only a small fraction of investment relative to those concepts. In light of this demonstrated performance, well-established scaling relations, and opportunities for further innovations, here we advocate for pulsed magnetic fusion as the most attractive path towards commercialization of fusion energy. 
\end{abstract}

\maketitle

\section{Introduction}\label{sec:intro}
 Increasing urgency for carbon-free energy sources \cite{energy_2023} and recent scientific advances in fusion \cite{ARC_2015, NIFLawson_2022, NIFTargetGain_2024} have catalyzed a national conversation regarding the most promising path to commercial fusion energy. 
  
Pulsed magnetic fusion (PMF), in which a rapid magnetically-driven implosion reaches fusion conditions, spans significant ranges in $P \tau$ phase space (e.g. $\tau$ from O(100ps) to O($\mu$s) with a corresponding $P$ to exceed Lawson's criterion). In recent years, leading PMF platforms have demonstrated $P \tau$ performance exceeding the records achieved with steady-state magnetic confinement devices (e.g., tokamaks) and on par with laser-driven inertial confinement fusion (ICF) experiments at similar facility scale (e.g., OMEGA) \cite{Wurzel_2021, knapp2022Bayesian} (as the only ignition-scale ICF facility, the NIF has the highest measured $P \tau$). Despite this performance, PMF remains less explored than magnetic confinement and laser-based ICF approaches.

\begin{figure*}
\centering
\includegraphics[width=5in]{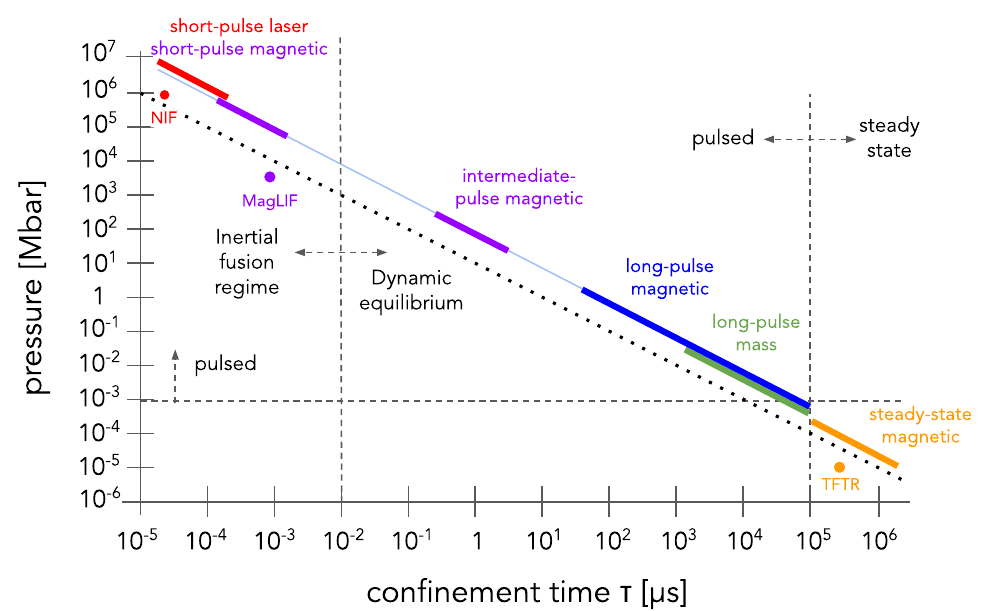}
\caption{\label{fig:SfA}  Parameter space of pressure ($P$) and confinement time ($\tau$) for fusion approaches, adapted from Science for America\cite{SfA}. Pulsed magnetic fusion can operate across the regimes shown in purple.}
\end{figure*}

As the U.S. moves towards a fusion-powered future under the White House’s ``bold decadal vision,''\cite{BoldDecadal} we believe PMF to be the most attractive path forward when balancing technology maturation, cost, and complexity. We find support for this conclusion in a recent JASON study \cite{JASON}, which considered a broad parameter space including PMF as a low-cost path towards fusion energy, and in a white paper on fusion by the Science for America organization \cite{SfA} (see Fig. \ref{fig:SfA}).

\noindent Our PMF community is driven by the following key principles:
\begin{itemize}
\item{The world is in immediate need of clean, deployable energy solutions including fusion.}
\item{Pulsed magnetic fusion is the best way to achieve high gain and high yield fusion, relevant for energy generation, and it has the potential to operate at lower stored energy and to be significantly more compact than competing technologies.}
\item{The U.S. is the leader in pulsed power technology and is in ideal position to lead high gain and high yield pulsed magnetic fusion, for both energy applications and national security, respectively.}
\end{itemize}

\noindent Our key principles imply that to meet the bold decadal vision through PMF, we must do the following:
\begin{itemize}
\item{We must rapidly advance reliable, high power pulsers capable of high repetition-rate.}
\item{We need a paradigm shift in pulser operational excellence to drive experimental innovation.}
\item{We must advance high $P \tau$ target designs in multiple pressure regimes.}
\item{We need target design codes and power flow modeling tools that are publicly available and have a large user base.}
\item{We need state-of-the-art diagnostics for inferring $P \tau$ and critical plasma conditions that cover a wide range of parameter space and capable of rep rate.}
\item{We need to form partnerships across the fusion energy ecosystem to advance materials science, tritium processing, and commercial engineering code development for fusion power systems.}
\item{We need a focused science and engineering outreach program (ZNetUS and others) to develop future leaders and the supply chain that spans industry, national laboratories, and academia.}
\item{By the end of the decade, we can and must demonstrate facility gain ($Q_f>1$) and remove significant technology hurdles to commercialization.}
\end{itemize}

\section{High-level vision}
\label{sec:hlv}

To realize the U.S. bold decadal vision for fusion energy through PMF and to realize the potential of PMF across the spectrum of energy and national security applications, we believe that the following three facilities are required: %
\begin{enumerate}
\item{Demonstrate $Q_f>1$ by the end of decade.} %
\item{Demonstrate a fusion pilot plant within 5 years after $Q_f>1$ is demonstrated.} %
\item{Develop a high-yield source for NNSA's national security mission \cite{NNSA}.} %
\end{enumerate}
The first two facilities will advance the frontier of fusion energy. The first goal, to demonstrate $Q_f > 1$, will require producing $\sim 100$ MJ fusion yields from a ``single shot'' scientific PMF facility. The second goal, a commercially-relevant demonstrator, will likely be $100$s of MJ yields at fusion energy system-relevant shot rates and with power plant technologies ({\it e.g.} tritium breeding blanket, heat exchanger, load recycling, etc.).

The third facility has distinct requirements derived from NNSA's national security mission, identified in recent roadmaps as a multi-mission\footnote{including x-ray sources, dynamic material properties, and high-energy-density physics studies in addition to the capability to execute high yield fusion experiments} next-generation pulsed power (NGPP) capability\cite{sinars2020review}. It is possible that the NNSA’s most critical requirements could be met by the $Q_f>1$ facility, motivating the development of a public-private partnership during its planning phase to ensure the national security elements required of NGPP were included during the initial build.

While the facilities required to achieve these three ambitious goals are distinct, successful execution of any of the facilities will require addressing a common set of scientific and engineering considerations. Recent workshops on the basic research needs for inertial fusion energy\cite{BRN_2023} did not fully explore the opportunities in pulsed magnetic fusion; here we provide a community-driven look at the key components of a program to advance the technology readiness level for pulsed magnetic fusion energy (PMFE). In the following sections, we present these considerations together with technical opportunities for innovation.

\section{Pulser architectures}
\subsection{Overview}
The Z Facility at Sandia National Laboratories \cite{sinars2020review} is presently the world's largest and most powerful pulser.  Z represents the culmination of \textit{conventional} pulsed-power technology development.  The pulser architecture and components upon which Z is based were developed in the 1900s.

Z is driven by conventional Marx generators that produce a power pulse with a 1-$\mu$s rise time.  Since many physics targets of interest require a 100-ns pulse, Z includes multiple stages of electrical-pulse compression.  Such stages complicate the design, operation, and maintenance of Z, and reduce Z's electrical-energy efficiency.

An advanced PMFE pulser architecture has the following objectives: high energy efficiency, simplicity of design, economies of scale, and engineered safety. Over the past decade, pulser architectures have evolved towards these objectives by using single-stage electrical pulse compression, low-voltage switching, and impedance matching \cite{stygar2007architecture,stygar2015conceptual,stygar2017conceptual}. Here, we propose future PMFE pulsers use a simplified architecture that meets all of these objectives \cite{stygar2017impedance}.

\begin{figure*}
    \centering
    \includegraphics[width=5in]{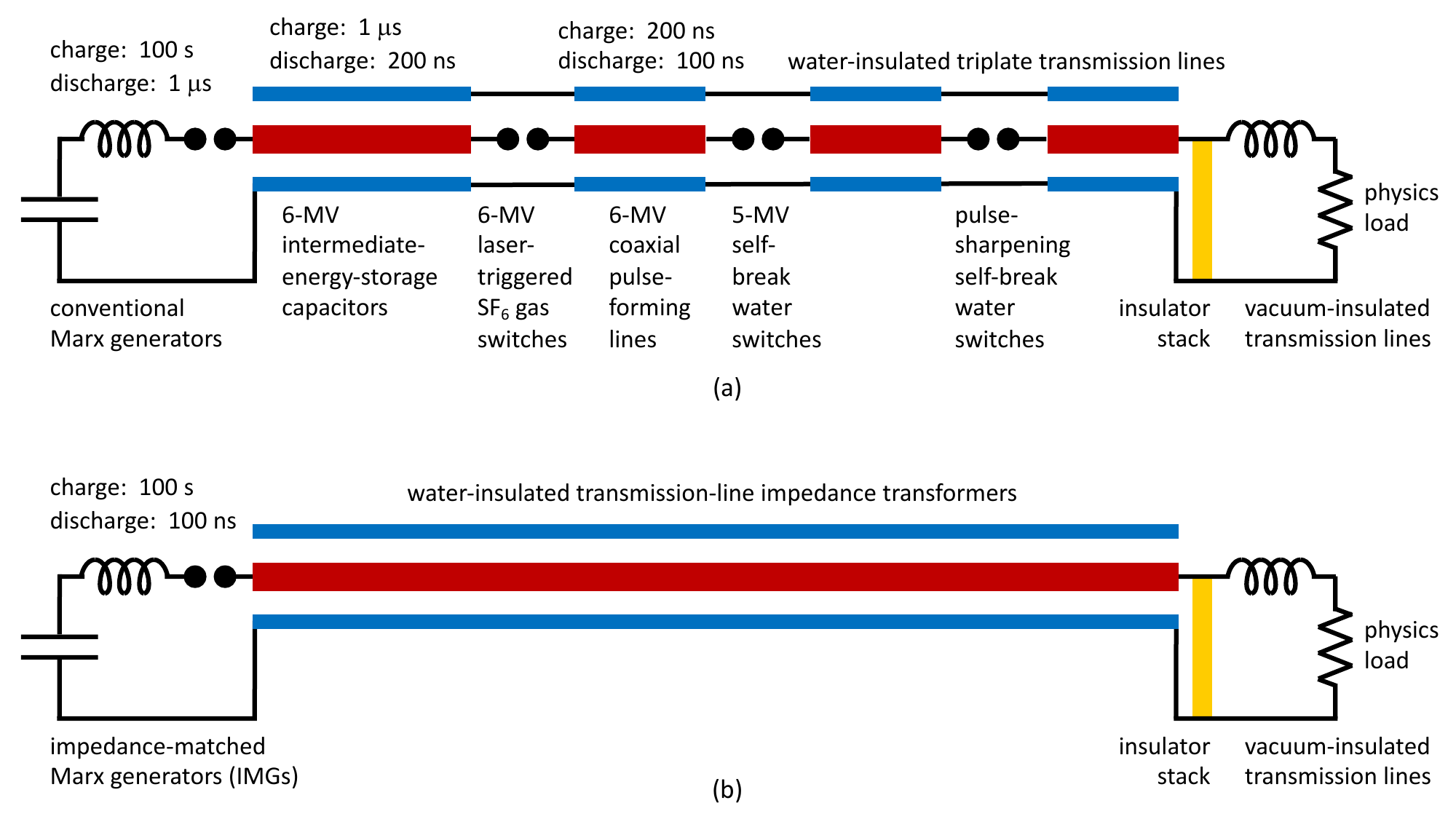}
    \caption{(a) Conventional pulser architecture.  (b) Advanced pulser architecture, which eliminates pulse compression and achieves direct and efficient delivery of 100 ns discharges from capacitors to the load.}
    \label{fig:architecture}
\end{figure*}

Figure \ref{fig:architecture} compares the conventional and an advanced pulser architecture known as an \textit{impedance-matched} Marx generator (IMG). Figure \ref{fig:architecture} illustrates the simplicity of the advanced concept, which goes directly from DC charged capacitors to a 100-ns pulse transmitted to the load in a single step.

\subsection{Advantages of the advanced architecture}

The conventional Marx generator was invented a century ago, in 1924, by Erwin Marx.  Since then, most pulsers have been powered by such technology.  A Marx is an electrical circuit that charges capacitors in parallel, and discharges them in series.  A conventional Marx is modeled with 0-D lumped circuit elements, and is designed to stack \textit{voltages}.  

The IMG concept was invented in 2017 \cite{stygar2017impedance}.  Like a conventional Marx, an IMG also charges capacitors in parallel and discharges them in series.  However, an IMG is designed to account for electromagnetic-wave propagation:  an IMG is modeled with 1-D transmission-line-circuit elements and is designed to stack \textit{waves}.

\begin{figure}
    \centering
    \includegraphics[width=5in]{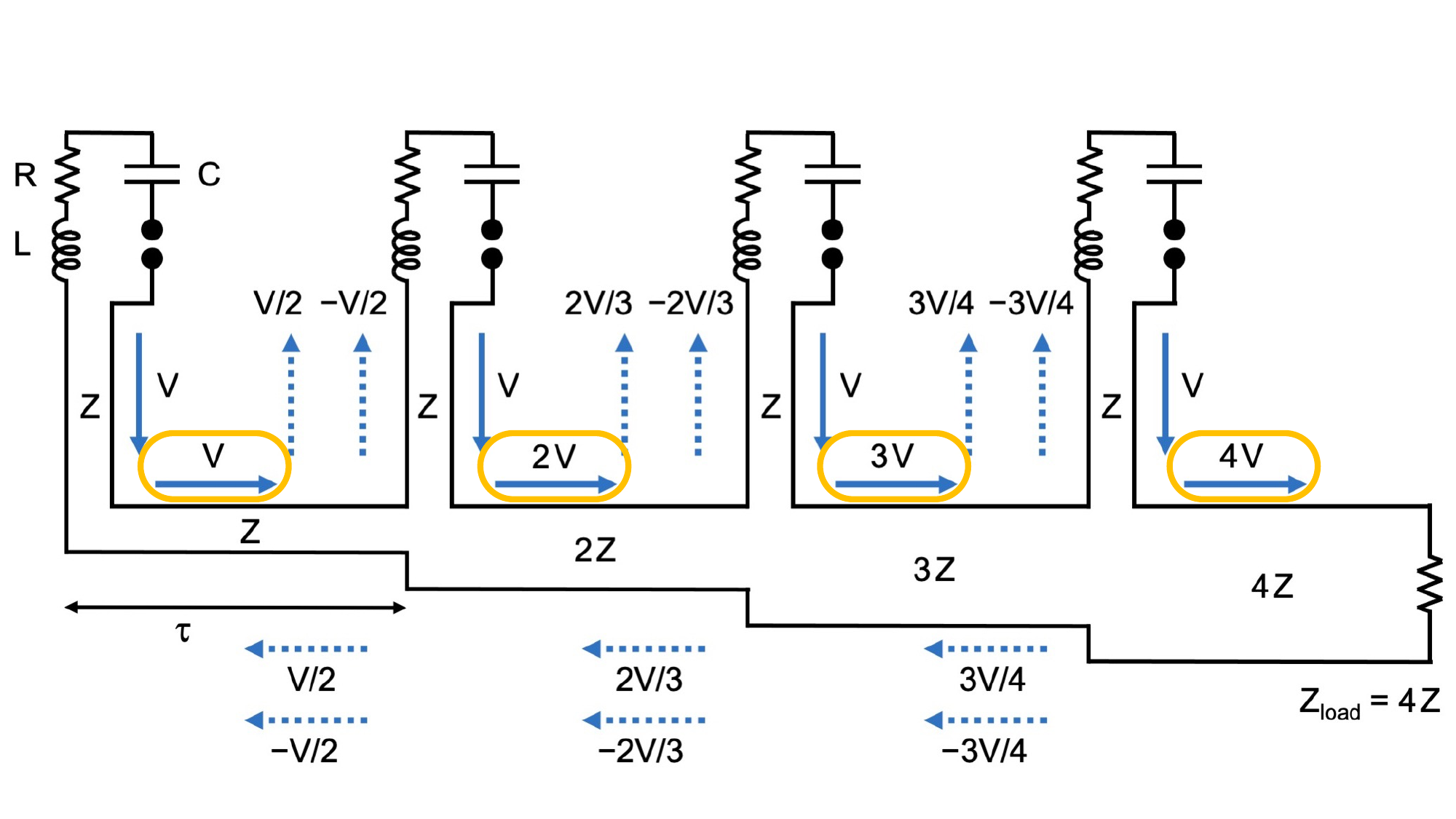}
    \caption{Transmission-line-circuit model of a 4-stage IMG.  Whereas a conventional Marx is modeled with 0D lumped circuit elements and is designed to stack \textit{voltages}, an IMG is modeled with 1D transmission-line-circuit elements and is designed to stack \textit{waves}.  By design, all the internal reflected waves within an IMG cancel; only the forward-going waves remain.}
    \label{fig:IMG}
\end{figure}

Figure \ref{fig:IMG} illustrates a transmission-line-circuit model of a four-stage IMG.  An IMG is a pulsed-power analog of a laser, with an energy efficiency of 90\%. (For comparison the flashlamp-pumped NIF laser is $<1\%$ efficient, and modern diode-pumped lasers are $\sim 10\%$ efficient).

\begin{figure}[h]
    \centering
    \begin{subtable}[b]{0.48\textwidth}
        \centering
        \includegraphics[trim=0 0 0 0.in, clip,width=1\linewidth]{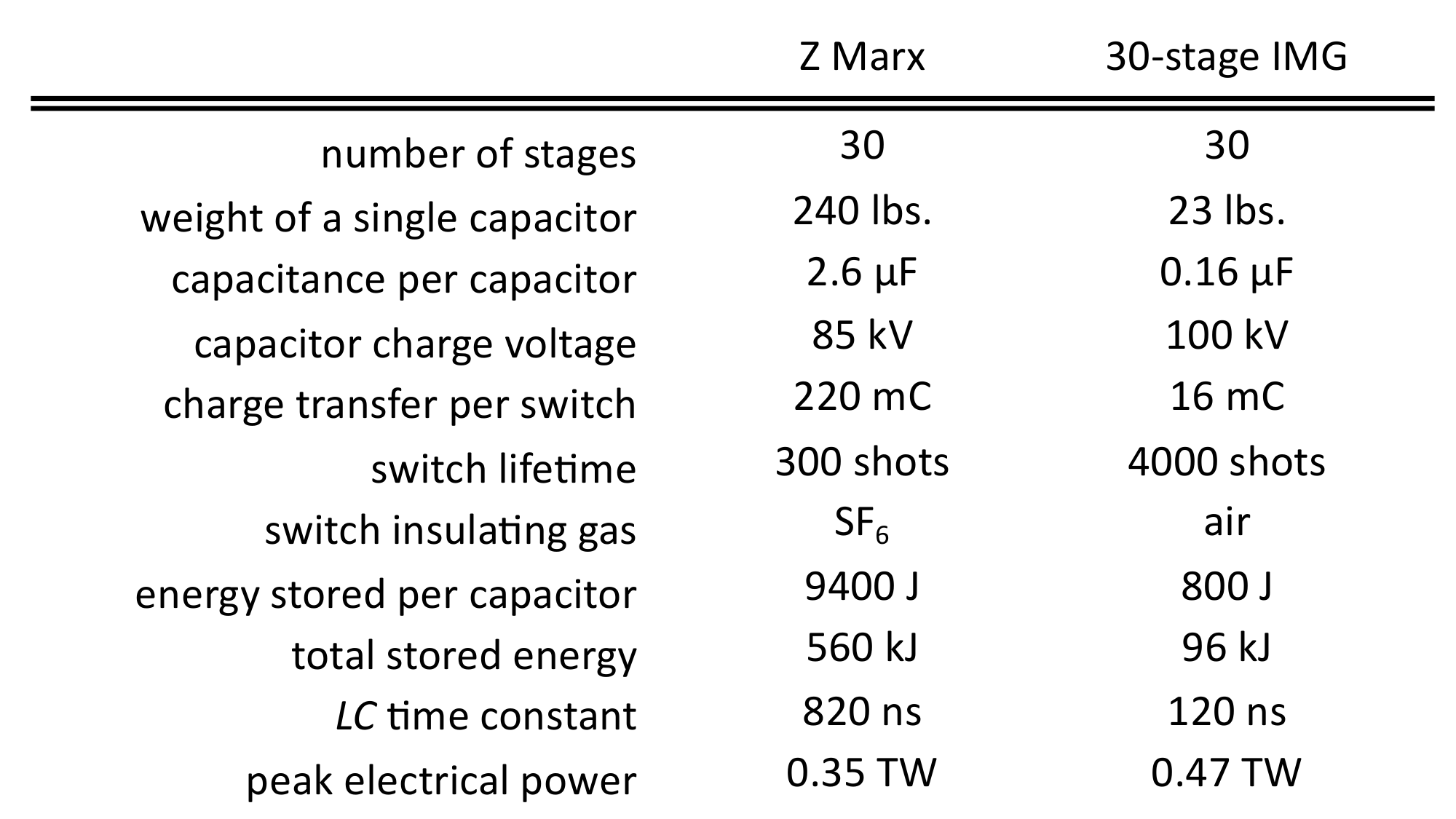}
        \caption{Parameters of a 30-stage Z Marx generator (which is a conventional Marx) and 30-stage IMG. \\}
      \label{fig:30stage}
\end{subtable}
    \hfill
    \begin{subfigure}[b]{0.45\textwidth}
     \centering
     \includegraphics[trim=0 0 0 0in, clip, width=1\linewidth]{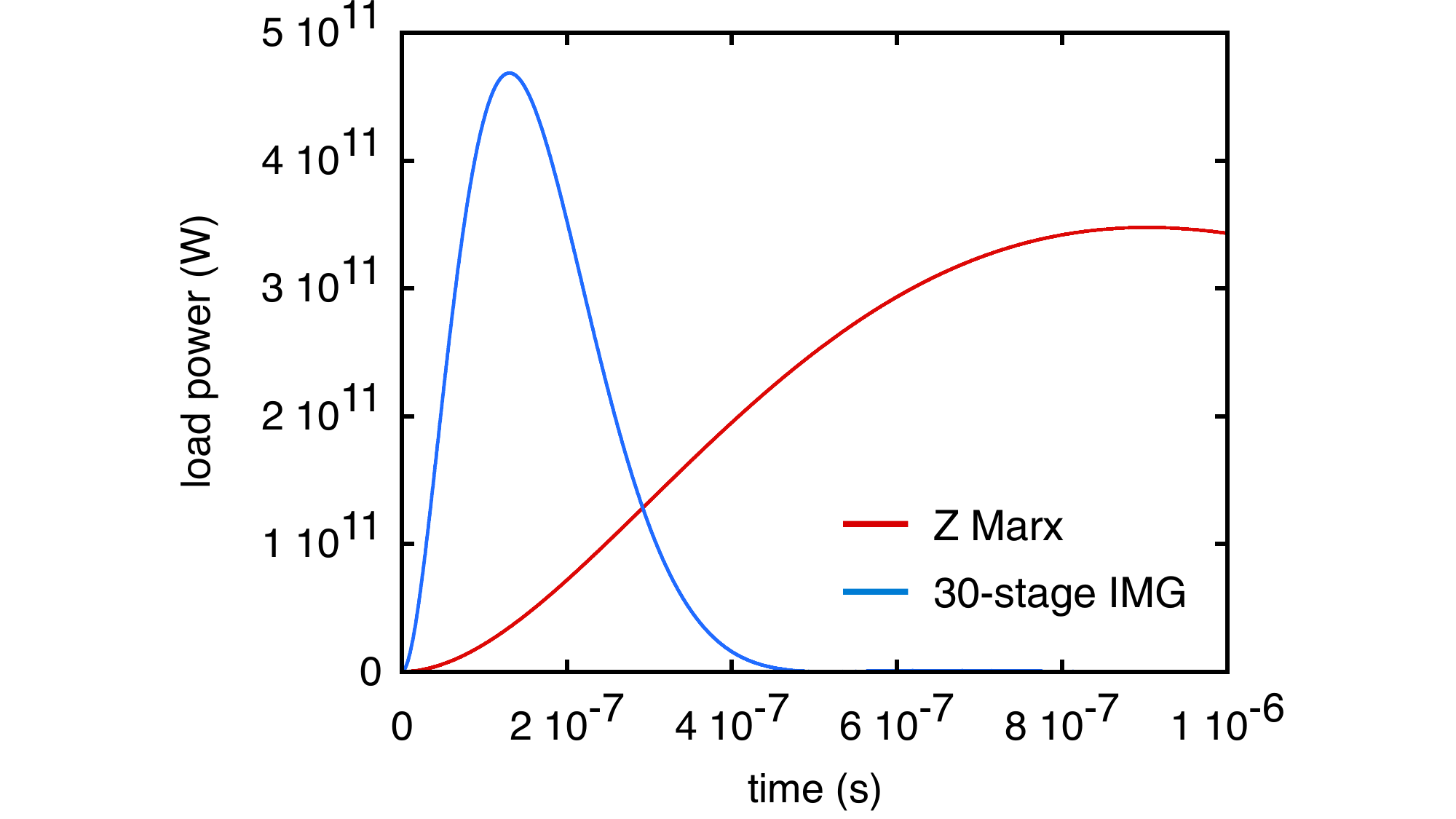}
     \caption{Electromagnetic-power time histories of a 30-stage Z Marx and 30-stage IMG. The IMG stores a factor of 6 less energy, and generates 34 percent more power.}
     \label{fig:power}
    \end{subfigure}
    \caption{Parameter comparison and power-time histories of conventional and Impedance-Matched Marx generators}
\end{figure}

Table \ref{fig:30stage} compares parameters of a Z Marx generator (which is a conventional 30-stage Marx) with those of a 30-stage IMG.  As described by the table, even though the IMG stores a factor of 6 less energy, it generates 34 percent more power. Figure \ref{fig:power} compares the time history of the electromagnetic power generated by the 30-stage Z Marx with that of the 30-stage IMG. Today, implementation of the architecture in practice has been limited to the 4-stage, 8-brick 60-GW Sirius-1 prototype \cite{lechien2023sirius}. %

By design, the advanced pulser architecture delivers the following:
    
\subsubsection{Increased energy efficiency}
Each Z Marx generates a power pulse with 1-$\mu$s rise time.  Since many physics targets of interest must be driven by a 100-ns pulse, the Z machine requires four stages of electrical pulse compression.  By design, the IMGs of the new architecture directly generate a 100-ns pulse. Because electrical pulse compression reduces efficiency, the calculated energy efficiency of the new architecture is twice that of the Z machine\cite{lechien2023sirius}.
    
\subsubsection{Increased component reliability and lifetimes}
The Z Facility is powered by switches that operate at voltages as high as 6 MV.  All the switches of the new architecture operate at 200 kV; lower-voltage switches are easier to design, fabricate, assemble, operate, trigger, and maintain. As indicated by Fig. 3, the switches of the Z Marx generators transfer 220 mC per shot; IMG switches transfer only 16 mC. 
Lower charge transfer will lead to longer operational lifetime and reduced maintenance. These improvements are expected to be significant, approaching an order of magnitude.

\subsubsection{Reduced pulser cost}

A pulser that delivers a given current to a physics target has a reduced stored energy with the new architecture compared to a conventional system because of its improved energy efficiency. The physical size is also reduced by approximately a factor of 2 in area.
These factors lead us to expect that the cost of a new architecture pulser is significantly less than a conventional architecture. Additional cost savings will result from the immense simplifications offered by the new architecture.  More specifically, the new architecture eliminates the following components of existing conventional pulsers: SF$_6$ recirculation and processing, $> 200$-kV switching and energy storage, laser triggering systems, and high voltage water switching.

\subsubsection{Commercialization}

In the advanced pulser architecture, the basic building blocks of energy storage and switching components enable Hertz-scale repetition rates and significantly improved reliability for long-life applications such as fusion energy. This is because the RC charging time constant of the distributed architecture of the IMG is 20 times lower than conventional pulser technology, while the coulomb transfer is lower by no less than an order of magnitude (see Table \ref{fig:30stage}. Moreover, component optimization takes place on a relatively small physical scale (e.g. individual capacitors and switches). Advanced pulser architectures like the IMG thus open a path to mass-manufacturing that is not otherwise open to conventional pulser architectures.

Gaps in technology exist that need to be bridged to achieve the durability, high rep-rate, and affordability necessary for the commercialization of fusion energy, regardless of the system's design. For instance, lasers, tokamaks, and pulsers all fundamentally rely on pulsed power systems. Regardless of fusion approach, to meet the capital expenditure (CapEx) goals for commercial viability, technology improvements are needed. For example, the energy storage and switching component replacement lifespan must extend by at least a factor of 1000 at Hertz operating rate. The cost of energy storage and switching must decrease by a factor of 5 to 10. There are design optimization axes to reduce coulomb transfer by several orders of magnitude in advanced pulsers to meet commercialization targets.

\subsection{Advancing Technology Readiness Level}
To advance pulser technology readiness level (TRL) at reactor-compatible energy, efficiency, and repetition rate, pulsed magnetic fusion power systems require maturation of pulser architectures (such as IMGs) that are capable of Hertz repetition rate, multi-million shot lifetimes, and production scale mass manufacturing. Focused investment in high reliability energy storage (i.e. capacitors) and switching technologies capable of very high power (100 kV, 50+ kA) benefits all fusion approaches, including PMFE.

\section{Target physics, fabrication, and experimental capabilities}
\begin{figure*}
    \centering
    \includegraphics[width=5in]{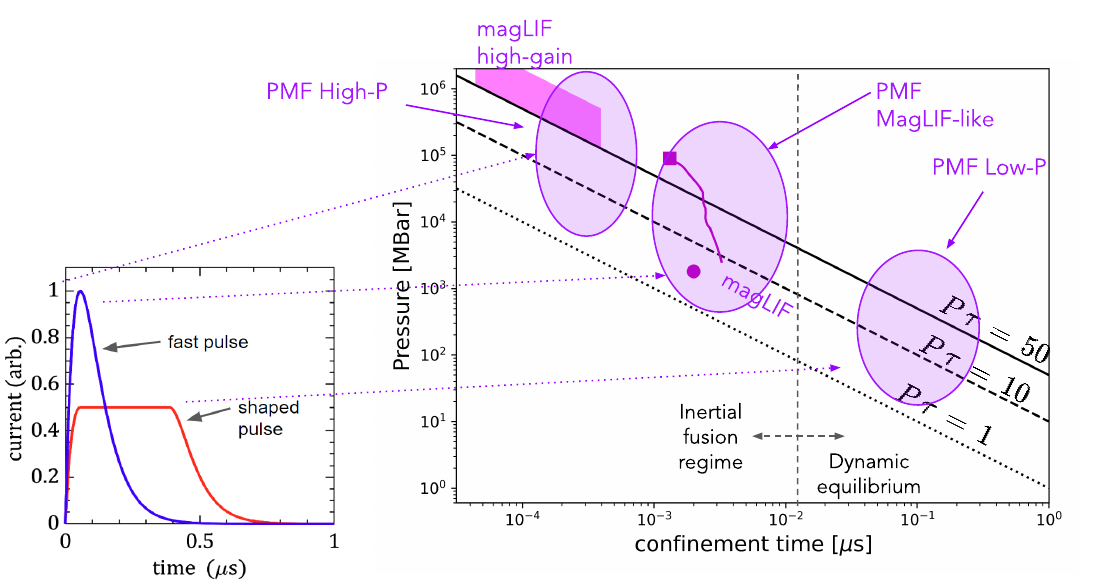}
    \caption{\label{fig:ptau} Schematic of the PMF parameter space for $P \tau$. Target classes from high to low pressure may be studied on a single pulser via pulse shaping and target design. Shown are data from Z \cite{knapp2022Bayesian} (purple point), a simulated current scaling gas-filled design \cite{Slutz2016Scaling} (purple curve and point), and estimated conditions for a high-gain $\sim$ GJ yield (shaded region, e.g. Ref. \citenum{Slutz2012HighGain,Slutz2016Scaling}).}
\end{figure*}
\subsection{Theoretical basis}

As mentioned in Section \ref{sec:intro}, a single PMF facility can explore target designs that operate over orders of magnitude in $P$ and $\tau$ respectively, as shown in Fig. \ref{fig:ptau}. We identify three regions in the $P\tau$ phase space, each with unique characteristics, all accessible using the same engineered pulser. At the center of this space is an inertial fusion regime where intermediate pressure systems can ignite when the fuel is highly magnetized (such as MagLIF, described later). Fig. \ref{fig:ptau} shows an experimental point from Z (purple circle) and a simulated scaling curve from 20 to 60 MA from Ref. \citenum{Slutz2016Scaling} for a gas-filled MagLIF target that produces up to $\sim 100$ MJ yield. To the left is an inertial fusion regime with higher pressure and lower confinement time (PMF high-$P$). Slutz and others have published MagLIF simulated designs with ice layers that produce in excess of GJ yields; a simple estimate of $P\tau$ for such shots is to assume similar liner $\rho R$, that the yield $\sim P^2 \tau$ at high ion temperature, and that hydro confinement $\tau \sim 1/P$ which produces the high-gain shaded region shown. To the right is a dynamic equilibrium regime where $\sim \mu$s duration pulses to drive targets at lower pressure with longer confinement time (PMF low-$P$).

In the inertial regime, the requirements to heat and compress fusion fuel to ignition conditions are independent of the compression scheme. Lasers such as the NIF accelerate a DT fusion fuel shell to high velocity ($v \geq 370$ km/s) to compressively heat the central vapor hot spot. Such implosions are susceptible to hydrodynamic instabilities, but laser direct drive (LDD) and laser indirect drive (LID) implosions benefit from ablative stabilization. In pulsed magnetic fusion, magnetically driven shells (liners) are continuously driven throughout the power pulse. For a recent review of ICF theory see Ref. \citenum{hurricane2023physics}.

The target at the core of all pulsed DT fusion schemes must be hot enough to have high fusion reactivity (usually 10 keV temperatures or above) and be confined such that self-heating, from energy released in the form of 3.6 MeV alpha particles, compensates for or surpasses energy loss mechanisms. In the inertial fusion regime, the hot fuel's self-heating surpasses all loss mechanisms and ``ignites,'' initiating a thermonuclear explosion in which the burn is only quenched by hydrodynamic expansion of the fuel, with a time scale set by its inertia. 

As discussed above, these requirements can be met by systems with a wide range of energy density or fuel pressure as long as the product of pressure and confinement time is large enough. In the extensive PMFE regime shown in Figure \ref{fig:ptau}, the portion where burn duration and confinement time is less than O(10)s of nanoseconds, plasma conditions are achieved by rapid implosion of a fuel-containing target i.e. the ``inertial fusion'' regime. 

For longer burn $\tau$ time scales (i.e. $>100$ ns), but with the overall system still pulsed (current rise time $<\sim 0.1$ s), a fusion system must operate in a ``dynamic equilibrium'' configuration, where the fuel is hydrodynamically quasi-static during the burn duration and the energy balance in the fuel is a competition between self-heating, auxiliary heating, and other energy loss mechanisms (e.g. radiation and thermal conduction). Plasma confinement is always required, with additional hydrodynamic confinement required when the plasma pressure exceeds the strength of materials (O(10) kBar).

PMF is the only fusion approach where a single facility can explore target concepts ranging from a classic inertial fusion regime, where the ignition physics has been demonstrated at NIF, to the dynamic equilibrium regime, where target physics is less explored but pressure requirements are more forgiving. This is illustrated by the wide range in confinement time shown in Figs. \ref{fig:SfA} and \ref{fig:ptau}. In all cases considered, the target physics time scales are orders of magnitude shorter than the time over which the chamber absorbs the energy emitted and resets its conditions for the subsequent shot.

\begin{figure}
    \centering
    \includegraphics[width=0.5\linewidth]{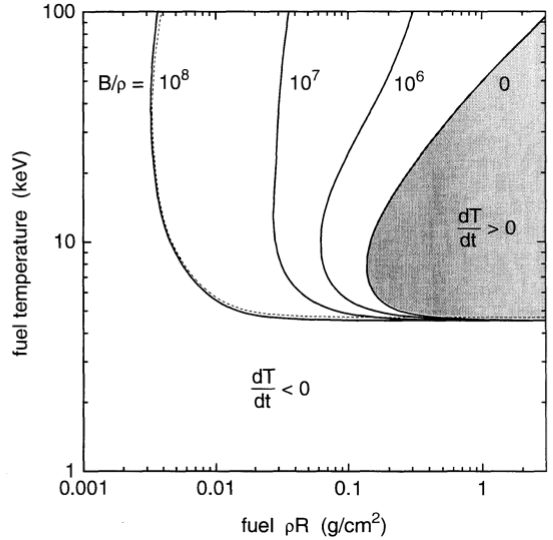} %
    \caption{Lindl-Widner diagram for magnetized DT cylinders at stagnation. The shaded area is the ICF ignition domain for $B=0$, similar to the region accessed by an igniting spherical NIF capsule. Curves to the left of this boundary show how a magnetic field expands the ignition space to lower $\rho R$. The curves are calculated for fixed values of $B/\rho$ in units of G cm$^2$/g. (from Ref. \citenum{Basko2000}).}
    \label{fig:BaskoRhoRT}
\end{figure}

One pulsed magnetic fusion concept in the inertial fusion regime is Magnetized Liner Inertial Fusion, or MagLIF; of the space shown in Fig. \ref{fig:ptau}, we focus only on MagLIF here as a well-documented, high-performing concept. In the MagLIF concept, a cylindrical metal liner is accelerated by driving a large current across its outer surface. The shell is driven at lower velocity $v \approx 100$ km/s to mitigate instabilities. Pre-heat energy and magnetic field, to suppress thermal conduction losses, are injected into the fuel cavity before the implosion stagnates, preconditioning the hot spot to ignite at lower stagnation pressure than a laser-driven implosion. MagLIF has a sound theoretical basis, extending the theory of Inertial Confinement Fusion (ICF) to include magnetized fuel \cite{Lindl1995,Basko2000}. Fig. \ref{fig:BaskoRhoRT} shows the $\rho R - T$ diagram for cylindrical magnetized hot-spots, with magnetization relaxing the requirement on fuel $\rho R$. Thus, MagLIF provides a magnetic direct drive pathway to ignition and gain that takes advantage of low-cost, compact pulsers.

\subsection{MagLIF scaling}

The MagLIF concept has been explored experimentally at the Sandia Z facility, using the same measurement techniques employed on large lasers like the NIF and Omega. Bayesian inference techniques are used to determine proximity to ignition via the generalized Lawson parameter $\chi = P \tau / P \tau_{ign}$, the ratio of the measured $P \tau$ to the value required for ignition. On Z, $\chi = 0.084 \pm 0.009$ has been demonstrated\cite{knapp2022Bayesian}. While Z is far from ignition, $\chi$ scales favorably with facility size for pulsed magnetic approaches compared to laser schemes. Conservative similarity scaling theory and numerical simulations give $\chi \propto I_{max}^3$, where $I_{max}$ is the peak drive current delivered to a target\cite{Ruiz2023Iscaling}. Facility size scales with delivered energy $E \propto I_{max}^2$, so $\chi \propto E^{1.5}$. In contrast, hydrodynamic scaling of laser-driven targets gives $\chi \propto E^{1/3}$. Thus, while the $\chi =  0.18$ obtained via laser direct drive on Omega is closer to ignition than Z, scaling this to a (symmetric) NIF-caliber 2.15 MJ laser results in $\chi=0.86$ and a fusion output of 1.6 MJ (target gain $< 1$)\cite{williams2024demonstration}. 
\begin{figure}
    \centering
  \begin{subfigure}[b]{0.48\textwidth}
    \includegraphics[width=1\linewidth]{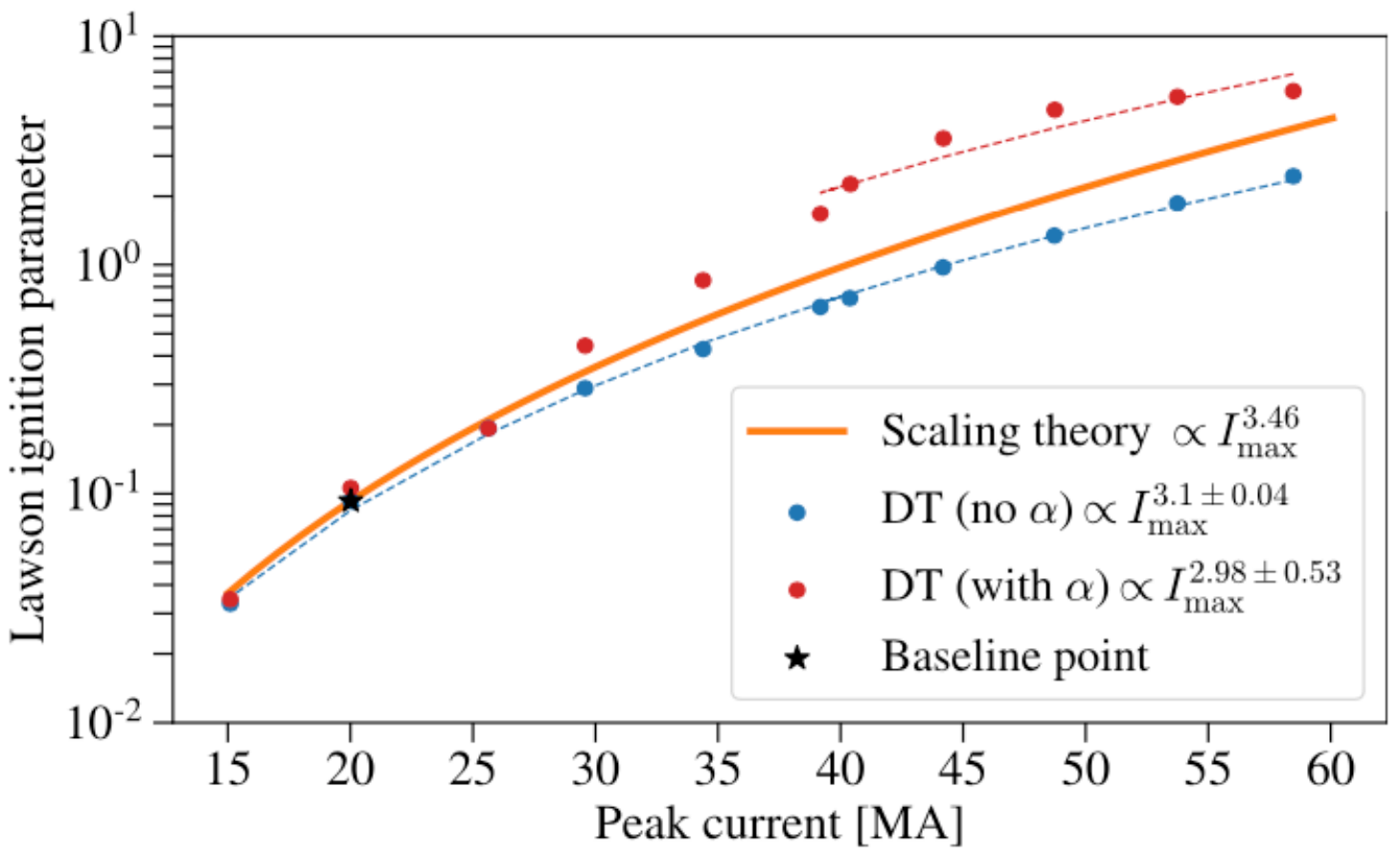}
    \caption{Lawson ignition parameter $\chi$ of  the similarity-scaled MagLIF loads. The sold line is from scaling theory. Dashed lines are power-law fits to 2D clean \textsc{hydra} simulations with and without $\alpha-$heating (solid circles).  (from Ref. \citenum{Ruiz2023Iscaling}).}
    \label{fig:RuizChi} 
    \vspace{12pt}
  \end{subfigure}
  \hfill
  \begin{subfigure}[b]{0.45\textwidth}
    \includegraphics[trim=0 0 0 0.1in, clip,width=0.97\linewidth]{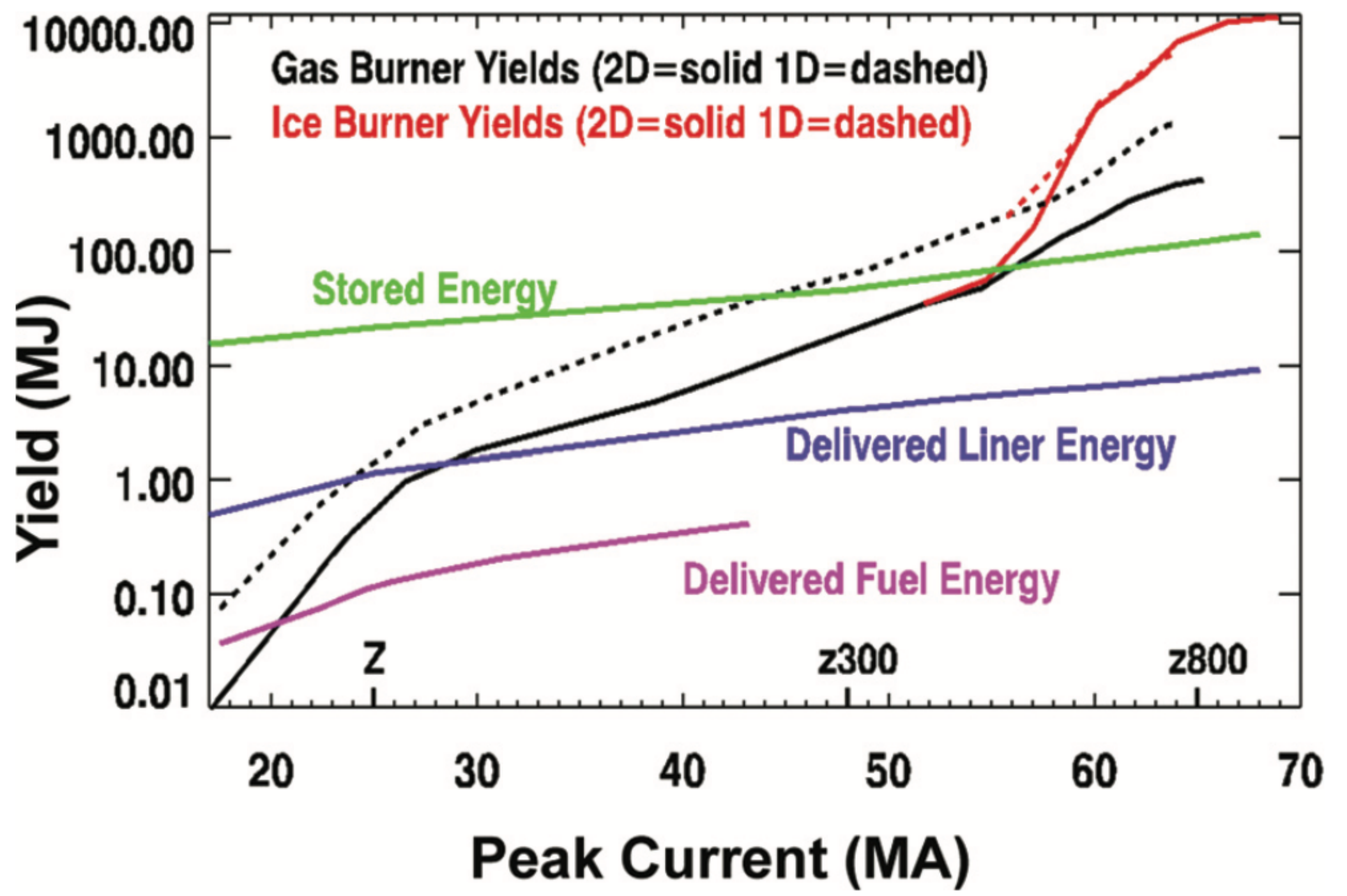}
    \caption{Yield vs. peak current for LASNEX simulations of MagLIF with (``Ice Burner'') and without (``Gas Burner'') a DT ice inner shell. For these designs, the 2D simulated yield jumps between 50 and 60 MA due to a thermonuclear burn wave propagating into the ice fuel. (from Ref. \citenum{Slutz2016Scaling})}
    \label{fig:SlutzYield}
  \end{subfigure}
  \caption{Ignition scaling curves for MagLIF loads as peak current is varied beyond the capabilities of the Z Facility}
\end{figure}
2D clean simulations using the rad-MHD code HYDRA, shown in Fig. \ref{fig:RuizChi}, show that $\chi \approx 2.4$ can be achieved by MagLIF driven at $I_{max} = 60$ MA, with simulated energy production $Y \approx 60$ MJ\cite{Ruiz2023Iscaling}. This design scaling is conservative, meaning it maintains the dynamics of implosions tested on Z by conserving the values of dimensionless parameters that describe the system (similarity scaling)\cite{SchmitRuiz2020,Ruiz2023Theory}. This suggests ignition with a cylindrical magnetized hot-spot is accessible to a reasonable pulser. Conservative scaling is presently being evaluated on the Z facility. Adding a cryogenic DT fuel liner to the implosion (not yet tested on Z) can mitigate impurity mix and increase the potential yield to many hundreds of megajoules\cite{Slutz2012HighGain,Slutz2016Scaling,RuizIFSA} (Fig.~\ref{fig:SlutzYield}). 

Thus, pulsed magnetic driven ICF is the shortest path to ignition that takes advantage of low-cost compact pulsers while retaining the ``ICF advantage'' --- the same facility can drive multiple designs and concepts, enabling rapid iteration and innovation. Pulsed magnetic targets are fabricated using the same technologies as laser targets, such as precision machining and electroplating. Pulsed magnetic targets are simpler than cryogenic hohlraum targets on the NIF, with fewer components and assembly steps, and with surface roughness requirements that are simpler to achieve - similar to that of 22-caliber bullet casings, which are made using rapid, low cost honing processes. 

Finally, we emphasize that while MagLIF is an ideal starting point, it is only the start: Innovations in target physics design and target fabrication will enable rapid progress for high $P \tau$ targets with higher and lower pressure than MagLIF. Experimental tests of innovative concepts are critical and are presently limited to publicly-funded facilities.

\subsection{Advancing Technology Readiness Level}
Demonstration of ignition and reactor-level gain requires a pulser that delivers multiple megajoules to an inertial fusion target, or equivalently delivering 50+ MA to a target region (r $<3$ cm). This facility will also enable innovation on target physics relevant for PMFE. Fabrication of reactor-compatible targets with cost-effective manufacturing methods at scale is required for all IFE approaches, including PMFE.

Access to public-sector PMF facilities such as the Z Facility, which are presently the only ones capable of testing integrated target physics concepts, would advance innovation in design in the near term.

\section{Simulations and modeling}

\subsection{Target Design}

Computational modeling has been essential to the advancement of fusion, including the achievement of laboratory ignition \cite{Kritcher_2024, Marinak_sims_ignition2024}. Designing high performance targets for future PMF facilities will require verified and validated radiation magnetohydrodynamics (MHD) codes available to a scientific community spanning national labs, academia, and private industry. Over the past decade, the PMF community has demonstrated the ability to model state-of-the-art PMF experiments to a high degree of accuracy \cite{Slutz_2018, Sinars_2010, Sinars_2011, Ruiz_2022, Awe_2016, Knapp_2017, Knapp_2020, Peterson_2013}. However, today the most heavily used and validated PMF design codes are not available for use outside of national laboratories and their chosen collaborators. Thus, a need exists for a more widely available, validated, rad-MHD design capability.

One exciting prospect for meeting the need for a widely used, verified and validated modeling capability is the FLASH code. FLASH \cite{Fryxell2000} is a publicly-available, parallel, multi-physics, adaptive mesh refinement (AMR), finite-volume Eulerian hydrodynamics and magneto-hydrodynamics (MHD) code, developed at the University of Rochester by the Flash Center for Computational Science (\protect\url{https://flash.rochester.edu}).
FLASH scales well to over 100,000 processors and uses a variety of parallelization techniques like domain decomposition, mesh replication, and threading, to optimally utilize hardware resources.
The FLASH code has a world-wide user base of more than 4,350 scientists, and more than 1,300 papers have been published using the code to model problems in a wide range of disciplines, including plasma astrophysics, combustion, fluid dynamics, high energy density physics (HEDP), and fusion energy.

Over the past decade and under the auspices of the U.S. DOE NNSA, the Flash Center has added in FLASH extensive HEDP and extended-MHD capabilities~\cite{Tzeferacos2015} that make it an ideal tool for the multi-physics modeling of PMF.
These include multiple state-of-the art hydrodynamic and MHD shock-capturing solvers~\cite{Lee2013}, three-temperature extensions~\cite{Tzeferacos2015} with anisotropic thermal conduction that utilizes high-fidelity magnetized heat transport coefficients~\cite{JiHeld2013}, heat exchange, multi-group radiation diffusion, state-of-the-art electrothermal transport coefficients~\cite{Davies2021}, tabulated multi-material EOS and opacities, laser energy deposition, circuit models, and numerous synthetic diagnostics~\cite{Tzeferacos2017}. 

\begin{figure}
    \centering
    \includegraphics[width=0.95\linewidth]{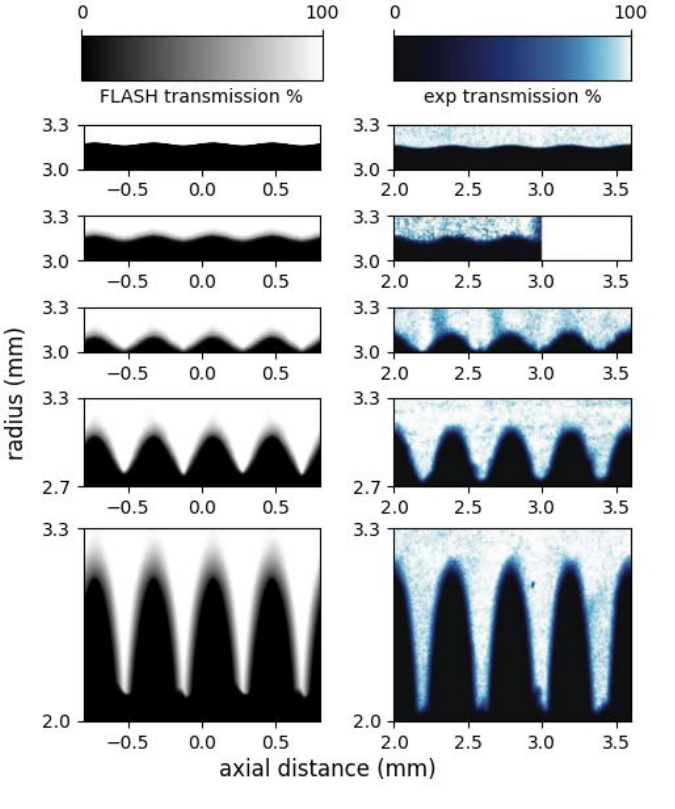}
    \caption{Synthetic radiographs of a FLASH simulation of an imploding aluminum liner with a pre-imposed 400 $\mu$m wavelength perturbation, compared to experimental radiographs from a series of magneto-Rayleigh Taylor benchmark experiments at Z described in Refs.~\cite{Sinars_2010, Sinars_2011}. From top to bottom, the images correspond to radiographs taken at 0, 42.7, 57.0, 67.7 and 83.0 ns.}
    \label{fig:flash_lincoln} 
\end{figure}

FLASH and its capabilities have been validated for over a decade through benchmarks and code-to-code comparisons~\cite{Fatenejad2013, Orban2022, Sauppe2023} and through direct application to numerous laser-driven plasma physics experiments~\cite{Falk2014,Yurchak2014, Meinecke2015, Li2016, Tzeferacos2018, Rigby2018, White2019, Meinecke2022}, leading to innovative science and publications in high-impact journals.
For pulsed-power experiments, FLASH has been able to reproduce past analytical models \cite{Slutz2001}, is being applied in the modeling of capillary discharge plasmas \cite{Cook2020}, and is being validated against gas-puff experiments at CESZAR \cite{Conti2020PRAB} and canonical liner implosion experiments at Z. For instance, Fig.~\ref{fig:flash_lincoln} shows FLASH simulations of the magneto-Rayleigh-Taylor aluminum liner platform described in Refs.~\cite{Sinars_2010, Sinars_2011}.

Accelerating the maturity of FLASH or any other widely available PMF design code will require enhanced collaboration across the PMF research community. Such enhanced collaboration has been recommended in the JASON study on low cost fusion development: ``The National Laboratories should contribute their unclassified state-of-the-art simulation codes to collaborations with academic and commercial efforts, and support the training of qualified users" \cite{JASON}. As a starting point, we recommend the community pursue the following, more limited collaborations and process improvements to rapidly benefit PMF simulation efforts.

Tabular material models for equation of state and transport coefficients, including magnetic resistivity, thermal conductivity, and single- and multi-group opacities, are known to have significant effect on the accuracy of the simulation results. These tables are maintained and controlled across a variety of institutions with a variety of access requirements and restrictions. Streamlining the processes for granting access to these tables while appropriately protecting the data and controlling their use would enhance the experimental relevance of the growing pulsed magnetic fusion simulation research effort. 

Additionally, there exists an opportunity to accelerate open-code V\&V efforts via focused collaborations between subject matter experts across the community. Similar scientific communities faced with comparable logistical complications have pursued a type of shared access development model for simulation and modeling abilities. This quasi open-sourced model seems to have accelerated the community’s computational abilities, as well as streamlined code verification and validation (V\&V) \cite{HEP_Roadmap_2019} without duplicating efforts. We advocate enhanced avenues for collaboration on the following basic science research topics: %
\begin{itemize}
\item{Identification of fruitful benchmark and validation problems for MHD and PMF}
\item{Best practices for linear solver settings for implicit diffusion equations (magnetic diffusion, radiation diffusion, thermal conduction) including preconditioner and solver settings}
\item{Treatments of floors and ceilings for densities and temperatures (e.g., vacuum density thresholds for magnetic resistivity)}
\item{Nominal algorithms and best settings for flux limited diffusion in the context of radiation diffusion and electron thermal conduction}
\item{Differencing algorithms for contact discontinuities (vacuum/conductor interfaces) and AMR refinement boundaries}
\item{Multispecies and multimaterial mixture rules for transport coefficients and EOS quantities spanning disparate parameter regimes (cold, solid to warm dense matter to weakly coupled plasma)}
\item{Development and identification of self-consistent EOS and transport coefficient tables}
\end{itemize}
Active collaboration on these topics will accelerate PMF simulation research, both by disseminating existing knowledge and by activating a larger community of scientists to advance the state-of-the-art in modeling methodology. The high energy physics community has benefited from similar calls for enhanced knowledge transfer to leverage all available expertise in pursuit of mission-critical software advances \cite{HEP_Roadmap_2019, Couturier_2017}.

\subsection{Power Flow and Pulser Design}

\begin{figure}
    \centering
    \includegraphics[width=1 \linewidth]{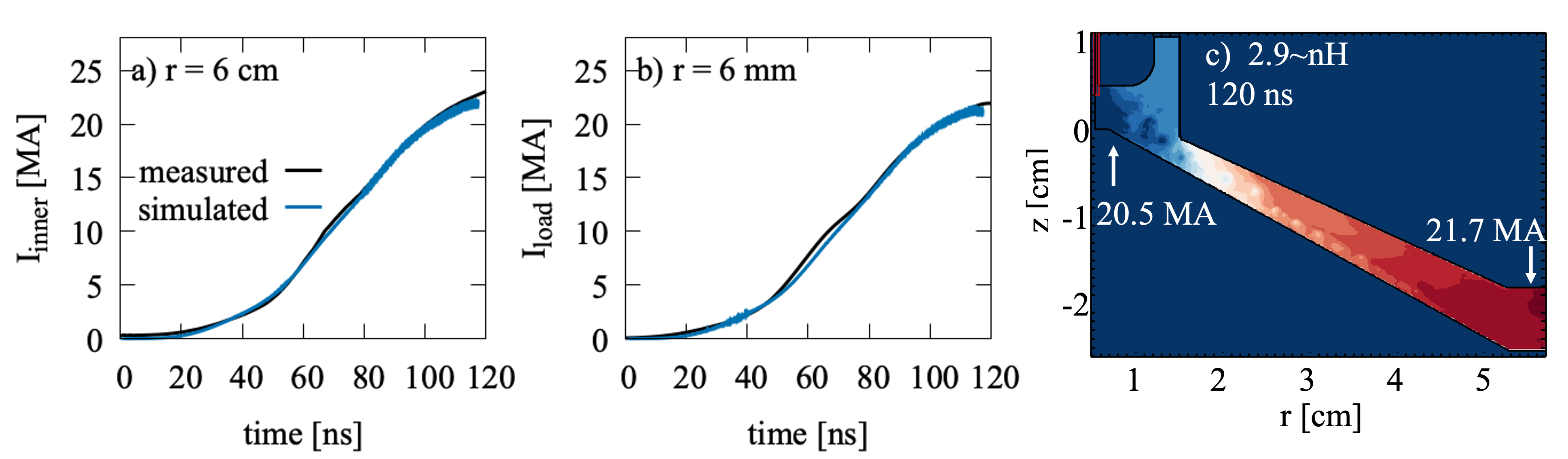}
    \caption{The measured and simulated current at a) 6 cm raduis and b) 6-mm radius are plotted. The contours of enclosed current are shown after 120 ns The data is from a low inductance (2.9 nH) Z shot. Figure from Ref. \cite{Bennett_milestone:2023}}
    \label{fig:Ben5-9} 
\end{figure}

Designing the PMF accelerator must include the impact of electrode plasmas created in the high power vacuum sections. The current delivered to the PMF load can be shunted by these plasmas. The state-of-the-art capability for the simulation of vacuum power flow in the accelerator has been demonstrated by the CHICAGO hybrid particle-in-cell (PIC) code.\cite{welch2019hybrid} CHICAGO permits high density and magnetic field plasma modeling with kinetic and/or fluid treatments\cite{welch2020hybrid}. The simulations include advanced surface physics and circuit modeling of upstream pulsed power components. Additionally, CHICAGO has demonstrated the ability to model the interaction of high-energy power-flow plasmas with PMF liners.\cite{tummel2022flow} CHICAGO capabilities now enable modeling of the heating of solid-density liners by power-flow plasmas within a single integrated simulation. The simulations have been validated with detailed comparisons of simulated current loss with Z data (see Fig. \ref{fig:Ben5-9}).\cite{Bennett2019transport}, \cite{Bennett2021transport} CHICAGO is currently being used to design next generation Z-pinch accelerators by LLNL, SNL and private companies.

CHICAGO is a general-purpose three-dimensional (3D) fully-electromagnetic PIC code designed for executing multi-scale plasma physics and pulsed power component simulations. Advanced field solver techniques combined with multiple plasma models (quasi-neutral, inertial-fluid and fully-kinetic) allow CHICAGO to treat large spatial and temporal scale problems. Additionally, CHICAGO has detailed material models for designing of accelerator components.\cite{thoma2023gaps} CHICAGO is commercially available through Voss Scientific (https://www.vosssci.com/products/chicago) with various options including one for universities.

\subsection{Advancing Technology Readiness Level}
The theory of inertial fusion is as applicable to pulsed magnetic fusion as it is to laser-based fusion, and simulation capabilities are comparable with validated radiation-hydrodynamic models having similar experiment-driven pedigree.

DOE/NNSA is the organization to establish appropriate controls, including export controls, for data and simulation codes for this application space.  We recommend that the PMFE community engage the NNSA laboratories to provide technical advice to DOE/NNSA on these controls. Agreements between national labs, industry, and academia that have been established in other domains may serve as a model. Similar to NIST data, publicly-funded unclassified tabular material data should be made publicly available when appropriate, with export controlled tables made available with appropriate restrictions. Cooperative Research and Development Agreements (CRADA) could be an appropriate venue for sharing models and data. Cost-sharing and a commitment to communicate model improvements to the laboratories could be made part of such agreements. 

\section{Diagnostics and measurement innovation}
\subsection{Overview}
High yield and high rep-rate pulsed magnetic fusion systems require advances in diagnostic and data analysis methodologies. The two classes of facility outlined in Sec. \ref{sec:hlv} place different requirements on diagnostics and analysis. A single shot gain demonstration facility requires diagnostic systems that enable inference of key physical parameters, e.g. $P\tau$, and identification of failure mechanisms,  allowing for the optimization of performance and gain. Such a facility will place a premium on diagnostic access during the design phase. In contrast, a rep-rate facility geared towards power generation will  require fewer physics diagnostics to understand the details of target operation and will be more focused on facility operation and health.  

\indent Because the first step will be the construction and operation of a gain demonstrator at relatively low rep-rate for the purposes of establishing target operation and optimizing performance, we must consider the associated diagnostic requirements and how they will impact facility design. Nuclear diagnostics are needed to diagnose burn history, burn volume, yield, and temperature.  Traditionally in ICF, x-ray diagnostics have proven valuable for understanding target performance, but x-rays require direct line of sight through vacuum to the target. This will inevitably impact the facility more than nuclear diagnostics, which can operate effectively in the presence of relatively thick, opaque windows. An additional consideration stems from the wide range of pressures and timescales accessible.  As shown in Figs. \ref{fig:SfA} and \ref{fig:ptau}, one facility and set of diagnostics may need to accommodate a wide range of pressures and burn durations that satisfy the $P\tau$ condition, impacting instrument design. At the higher pressure regimes, technologies implemented on NIF/Z/OMEGA or upcoming through the NNSA National Diagnostic Working Group effort may be adaptable to this problem, but innovation is required for longer burn durations. Chamber diagnostics must operate reliably in the presence of high neutron yields ($\sim 100$~MJ), which cause background radiation and activation of components and increase the risk of debris damage to electronics, optics, and other equipment.    

\indent Diagnostic access is a key design consideration for the pulser architecture, which affects the performance of the driver. These trade-offs with the facility design necessitate a quantitative means by which the value of an individual instrument can be weighed against its impact on the facility (performance, cost, schedule, etc.). This can be achieved through a system level model that integrates information from multiple diagnostics to make inferences about target performance. One can envision an optimization loop where diagnostics are included self-consistently with associated changes to the driver and the facility, target performance is simulated, and inferences are made from synthetic data. In this way, the information content of the instruments can be quantitatively weighed against their impact to the facility.  Establishing this framework also conveniently establishes the models and tools needed to rapidly analyze and integrate diagnostic data post-shot to form a coherent picture of target performance, allowing rapid iteration and progress.

\indent At high rep-rate, the diagnostic focus shifts from the target to the facility. System components must be monitored to ensure they are functioning properly. Monitoring the fusion chamber environment will also be critical so that an acceptable envelope of conditions can be maintained.  Additionally, \emph{in-situ} monitors for target tracking and other repetitive operational tasks are required. The real-time nature of these systems at high rep-rate invites the use of machine learning to allow the individual subsystems to adapt to changing inputs/outputs over time. For example, signals obtained from electrical components like capacitors and switches can be used to predict failure probabilities and flag items for replacement or isolation when probabilities grow unacceptably high.  Research should be started immediately so that useful systems are available by the time the community is ready to build a commercial pilot plant. 

\subsection{Advancing Technology Readiness Level}

Adaptation of existing diagnostic capabilities on flagship facilities (e.g. NIF, Z, OMEGA) will be key for developing next-generation PMFE facilities. Collaboration between national laboratories, universities, and private industry is needed to further advance diagnostic technologies and analysis methods. Research into diagnostic systems needed for a pulsed fusion pilot plant must begin immediately and has common challenges and TRL across IFE approaches; a new small-scale higher-repetition-rate pulser could benefit the PMF community specifically.

\section{Chamber design and engineering}
\subsection{Overview}
There are multiple facilities proposed in Section \ref{sec:hlv}.  They can be divided into two general classes. The first is single shot facilities for pulser and high gain platform development, national security or other missions that could benefit from intense radiation and fusion neutron fluxes.  Examples include nuclear waste processing and tritium production. The second are rep-rated facilities, particularly those aimed at power generation, that need to fire at rates approaching 1 Hz with high average power, i.e., $P_{avg} \geq$ 100 MW. While there are no insurmountable hurdles, there are multiple engineering challenges to pulsed fusion energy systems, including\cite{greenspan1986fusion}: 

\begin{itemize}
    \item a tritium-producing blanket to replenish burnt, lost, and decayed inventory
    \item a target chamber wall robust to high average power fusion product loading
    \item rapid high gain target production and injection
    \item rapid and sufficient clearing of the chamber to support continuous operations for up to 6 months
\end{itemize}

In the past few decades there have been fusion energy system studies that have addressed these issues for laser and ion beam driven fusion concepts.  These include SOLACE, HYLIFE-II \cite{moir1994hylife}, LIFE \cite{latkowski2011chamber}, Z-IFE \cite{meier2004analyses,meier2007systems}, and similar studies for single shot applications for X-1 \cite{peterson1998x}.  The most promising solutions devised in these studies will be reviewed using new computational tools and with modern material considerations for pulsed magnetic fusion systems.

For DT-based fusion systems, about 80\% of the fusion energy released streams out of the fusion plasma in the form of 14 MeV neutrons. These pass easily through target materials and walls/hardware and are absorbed in a surrounding blanket. For pulsed fusion systems, be they laser- or pulser-based, much of the remaining energy from the alpha particles is deposited locally and vaporizes the target. The vaporized target radiates and expands into the surrounding chamber, which must be capable of withstanding the resulting thermal and mechanical loading. 

In pulsed magnetic fusion systems the target is directly coupled to the pulser, which comes with two unique advantages. First, this reduces by two orders of magnitude the positioning requirements in rep-rate applications, i.e.,  $\sim$mm scale positioning requirements for pulsed magnetic fusion systems compared to $\sim10$-$\mu m$ positioning for NIF capsules. Second, the power flow hardware between the target and the vacuum chamber dielectric stack provides physical and optical protection of the insulators, thus avoiding the ``direct exposure" problem of laser-based optics systems. Inline pressure-pulse baffling is also possible, since pulsed electrical energy can be made to follow curvilinear surfaces. Low-mass, low-cost magnetically insulated transmission lines (MITLs) designed to fully sublimate during the rising pulse need to be developed. These can be tested on single-shot pulsers, informed by modeling/simulation tools developed for power flow studies.

For all pulsed fusion power systems including pulsed magnetic fusion, safety, handling of radioactive byproducts, and efficient coupling of heat exchangers and balance of plant optimization are challenges the community faces. Pulsed magnetic fusion systems are, however, inherently shielded by the $\sim$ million gallon water insulated power delivery systems that deliver power to the vacuum region.

\subsection{Advancing Technology Readiness Level}
Aspects of chamber design, and development of appropriate first-wall materials, are common across IFE approaches and investment is needed to advance the TRL. Target injection, tracking, and engagement at reactor-compatible specifications is a commonality for IFE; PMFE has the advantage of operating at a relatively low repetition rate (e.g. Hz) with the unique aspect of electrically coupling to the target at reduced alignment tolerances versus laser-based approaches. Also, the upper and lower axial extents of the fusion target chamber are open allowing for significant access to the target region for target insertion equipment (targets can be launched through free space like in some concepts for laser-based IFE, but they can also be inserted with mechanical positioning equipment that remain attached to the target during fusion operation with no degradation of fusion performance). Inertial fusion requires minimizing the mass of vaporized material - testing of revolutionary low-mass concepts for PMFE is needed at existing academic-scale and publicly-funded facilities such as the Z Facility. 

A key cost driver for techno-economic analysis is in cost scaling of key components, which can inform modern assessments of IFE concepts including PMF. Advanced cost optimization models and tools have been developed for tokamaks; investment in new reduced models is required so those tools may be applied to pulsed fusion systems to advance techno-economic analysis of reactor concepts.

\section{Ensuring Long-term Intellectual Leadership}

A critical factor in maintaining leadership in pulsed magnetic science and technology (PMS\&T), including the area of pulsed magnetic fusion energy, is a dedicated initiative to train a new specialized workforce. This includes fusion engineers, pulser architects, target physicists, experimentalists, computational physicists, and precision fabricators. The ZNetUS program \footnote{https://znetus.eng.ucsd.edu/} was launched in 2022 precisely for this purpose. ZNetUS represents a collaboration of experts from academic institutions, national laboratories, and the private sector, committed to advancing the fields of pulsed magnetic science, technology, and high-energy density physics for both energy and national security. The program's central goal is to cultivate a diverse cadre of future scientific leaders.

ZNetUS's mission encompasses the following objectives: i) organizing annual workshops, ii) managing a User Facilities Program, iii) coordinating a cross-institution transformational technologies development effort, and, iv) coordinating advanced code development activities to develop publicly accessible, high-impact simulation codes. 

\section{Conclusion and next steps}

Pulsed magnetic fusion must be a key component of the fusion landscape to realize the U.S. bold decadal vision for fusion energy, as we believe it represents the most attractive path towards commercialization. Here, we have articulated a set of community-developed principles and our own bold vision to develop three major advances: the demonstration of facility gain ($Q_f>1$) by the end of the decade, a subsequent commercial pilot plant, and a next-generation source for national security needs. Realizing this requires a vigorous program in the science and engineering of PMF. This program encompasses advanced pulser architectures, target physics, fabrication and experimental capabilities, simulations and modeling, diagnostic and measurement innovation, and energy system design and engineering. By building a community around this common vision for PMF and its supporting areas of science and engineering, we can achieve the bold vision laid out here. This document is therefore intended to advocate for PMF and begin organizing the community. 

\subsection{Summary of Recommendations}

Year over year since the early 1990s pulsed magnetic ICF has received a factor of 5 to 10 less investment than laser-based ICF and tokamaks. Nonetheless, pulsed magnetic fusion has demonstrated P$\tau$ performance comparable with both approaches at similar facility scale, presenting the classic ``innovator's dilemma'' \cite{christenson1997innovator}. 
Our view is that focused attention,
particularly to improving the distribution of resources within the publicly-funded fusion ecosystem, and focused investment to advance technology readiness level in the following areas will provide a similar rapid return on that investment.

\begin{itemize}
    \item Pulsed magnetic fusion power systems require maturation of pulser architectures that are capable of Hertz repetition rate, multi-million shot lifetimes, and production scale mass manufacturing.
    \item Focused investment in high reliability energy storage and switching technologies capable of very high power (100 kV, 50+ kA) benefits all fusion approaches, including PMF.
    \item Collaboration between fusion industry and national laboratories on target physics and innovative concepts is key to achieving high gain fusion and advancing the science of PMF.
    \item Access to public-sector PMF facilities such as the Z Facility, which are presently the only ones capable of testing relevant-scale target physics concepts, would advance innovation. 
    \item DOE/NNSA is the organization to establish appropriate controls, including export controls, for data and simulation codes for this application space. Agreements between national labs, industry, and academia that have been established in other domains may serve as a model.
    \item Similar to NIST data, publicly-funded unclassified tabular material data should be made publicly available, with export controlled tables made available with appropriate restrictions. CRADA could be an appropriate venue for exchanging models and data, potentially including cost-sharing and a commitment to share model improvements with the NNSA laboratories.
    \item National laboratories, universities, and private industry should collaborate to advance diagnostic technologies and analysis methods.
    \item Research into diagnostic systems needed for a pulsed fusion pilot plant must begin immediately and could include a new small-scale higher-repetition-rate pulser.
    \item Technological challenges in materials and chamber design have significant commonality across fusion approaches and advances are needed that will benefit PMFE.
    \item Inertial fusion requires minimizing the mass of vaporized material: testing of revolutionary low-mass concepts is needed at existing academic-scale and publicly-funded facilities such as the Z Facility.
    \item Cost optimization tools developed for tokamaks require investment in new reduced models so those tools may be applied to pulsed fusion systems.
    \item Academic-scale programs such as ZNetUS and publicly-funded facilities should be supported and made accessible to industry for long-term intellectual leadership and community growth.
\end{itemize}

\section{Acknowledgments}
We thank F. Beg, N. Nardelli, D. Rose, K. Peterson, and K. Raman, for review of the manuscript, and S. Davidson for contributions to Fig.~\ref{fig:flash_lincoln}.

This work was performed under the auspices of the U.S. Department of Energy by Lawrence Livermore National Laboratory under Contract DE-AC52-07NA27344. This article has been authored by an employee of National Technology \& Engineering Solutions of Sandia, LLC under Contract No. DE-NA0003525 with the U.S. Department of Energy (DOE). This paper describes objective technical results and analysis. The views and opinions expressed in this paper represents the individual views of the authors and do not necessarily represent the views of any of the affiliated national laboratories, U.S. Department of Energy or the United States Government. This work was supported by the U.S. Department of Energy through the Los Alamos National Laboratory. Los Alamos National Laboratory is operated by Triad National Security, LLC, for the National Nuclear Security Administration of U.S. Department of Energy (Contract No. 89233218CNA000001). The Flash Center for Computational Science acknowledges support by the U.S. Department of Energy National Nuclear Security Administration under Award Numbers DE-NA0003856, DE-NA0003842, DE-NA0004144, and DE-NA0004147, under subcontracts no. 536203 and 630138 with Los Alamos National Laboratory, and under subcontract B632670 with Lawrence Livermore National Laboratory. We also acknowledge support from the U.S. Department of Energy  Advanced Research Projects Agency-Energy  under Award Number DE-AR0001272 and the U.S. Department of Energy Office of Science under Award Number DE-SC0023246.

\bibliographystyle{unsrt}
\bibliography{main}

\end{document}